\documentclass{eas}
\usepackage{graphicx}
\begin{document}

\title{
Tracing star formation in galaxies with molecular line and continuum observations
} 
\author{Kohno, K.}\address{Institute of Astronomy, The University of Tokyo;
\email{kkohno@ioa.s.u-tokyo.ac.jp}}
\author{Muraoka, K.}\sameaddress{1}
\author{Hatsukade, B.}\sameaddress{1}
\author{Tanaka, K.}\sameaddress{1}
\author{Iono, D.}\sameaddress{1}
\author{Nakanishi, K.}\address{Nobeyama Radio Observatory, National Astronomical Observatory of Japan}
\author{Tosaki, T.}\sameaddress{2}
\author{Sawada, T.}\sameaddress{2}
\author{Kawabe, R.}\sameaddress{2}
\author{Ezawa, H.}\address{National Astronomical Observatory of Japan}
\author{Yamaguchi, N.}\sameaddress{3}
\author{Tamura, Y.}\sameaddress{1,3}
\author{Wilson, G.}\address{Astronomy Department, University of Massachusetts}
\author{Yun, M.S.}\sameaddress{4}
\author{Hughes, D.}\address{Instituto Nacional de Astrofisica, Optica y Electronica (INAOE)}
\author{Matsushita, S.}\address{Academia Sinica, Institute of Astronomy and Astrophysics (ASIAA)}
\author{Hsieh, P.Y.}\sameaddress{6}

\runningtitle{Kohno \etal: Tracing Star Formation in Galaxies}

\begin{abstract}

We report our recent progress on extragalactic spectroscopic and continuum observations,
including HCN($J$=1--0), HCO$^+$($J$=1--0), and CN($N$=1--0) imaging surveys 
of local Seyfert and starburst galaxies 
using the Nobeyama Millimeter Array,
high-$J$ CO observations ($J$=3--2 observations 
using the Atacama Submillimeter Telescope Experiment (ASTE) 
and $J$=2--1 observations with the Submillimeter Array) of galaxies, 
and $\lambda$ 1.1 mm continuum observations of high-$z$ violent starburst galaxies
using the bolometer camera AzTEC mounted on ASTE.

\end{abstract}
\maketitle
\section{Introduction}

Massive stars are formed in the densest regions of molecular clouds;
therefore, molecular spectroscopy of high-density tracers such as 
rotational lines of HCN and HCO$^+$ molecules in millimeter and submillimeter wavelengths
provides us with essential details on the origin of massive stars in galaxies.
They are free from dust extinction; hence, dust-enshrouded star formation 
can be unveiled using these observations.
These spectroscopic observations are also expected to be a crucial diagnostic of nuclear power sources of dusty active galaxies,
i.e., a probe to separately assess the nuclear star formation 
associated with active galactic nuclei (AGNs).
Both AGNs and starbursts will have a strong impact 
on the surrounding dense interstellar medium (ISM)
and form X-ray dominated regions (XDRs) and photo-dissociation regions (PDRs), respectively.
However, they will have drastically different physical and chemical
conditions that result in different molecular abundances and excitations 
between them (e.g., \cite{Maloney1996}; \cite{Meijerink2005}). 

In this paper, we report our recent progress on extragalactic spectroscopic and continuum observations,
including HCN($J$=1--0), HCO$^+$($J$=1--0), and CN($N$=1--0) imaging surveys 
of local Seyfert and starburst galaxies 
using the Nobeyama Millimeter Array (NMA),
high-$J$ CO observations ($J$=3--2 observations 
using the Atacama Submillimeter Telescope Experiment 
(ASTE; \cite{Ezawa2004}) 
and $J$=2--1 observations with the Submillimeter Array (SMA)) of galaxies, 
and $\lambda$ 1.1 mm continuum observations of high-$z$ violent starburst galaxies
using the bolometer camera AzTEC (\cite{Wilson2008}) mounted on ASTE.

\section{Tracing star formation toward active nuclei}

\subsection{HCN-enhanced nuclei (HENs): toward a new diagnostic probe of nuclear energy source}

The NMA imaging survey of HCN($J$=1--0) and HCO$^+$($J$=1--0) 
on local Seyfert and starburst
galaxies (NMA-DENSS project; Kohno \etal\ 2001, 2007, 2008a; Kohno 2005) 
revealed the presence of ``HCN-enhanced nuclei'' 
or HENs among some of the Seyfert galaxies (Fig.~1).
They show elevated HCN intensities with respect to CO and/or HCO$^+$ emissions;
i.e., HCN/CO integrated 
intensity ratios ($R_{\rm HCN/CO}$)$>$0.3 
and/or HCN/HCO$^+$ integrated intensity ratios ($R_{\rm HCN/HCO+}$)$>$1.5 
in the brightness temperature scale.
They are quite strange or intriguing because no clear evidence
of a nuclear starburst is shown despite the fact that HENs are very luminous 
in the HCN emission
(e.g., the lack of mid-infrared PAH emission, see table 1 of Kohno \etal\ 2007).
The absence of HENs among starburst galaxies 
and the detections of highly excited high-$J$ CO lines
in HENs (e.g., CO($J$=3--2)/CO($J$=1--0) integrated intensity ratio 
of $1.9 \pm 0.2$ at the center of M 51, \cite{Matsushita2004}; 
CO($J$=2--1)/CO($J$=1--0) of $1.9 \pm 0.2$ at the center of NGC 1097, Fig.~2) 
could also support the hypothesis that the elevated HCN($J$=1--0) emission is 
closely connected to the presence of 
X-ray irradiated dense molecular medium or XDRs.
This idea is successfully applied to local LIRGs/ULIRGs (e.g., \cite{Imanishi2007};
\cite{Garcia-burillo2007}), although further studies
will be required to understand the gap between the observations and theories
(e.g., \cite{Yamada2007}).

\begin{figure}
   \begin{minipage}{.4\textwidth}
    \hfill 
\includegraphics[width=7cm]{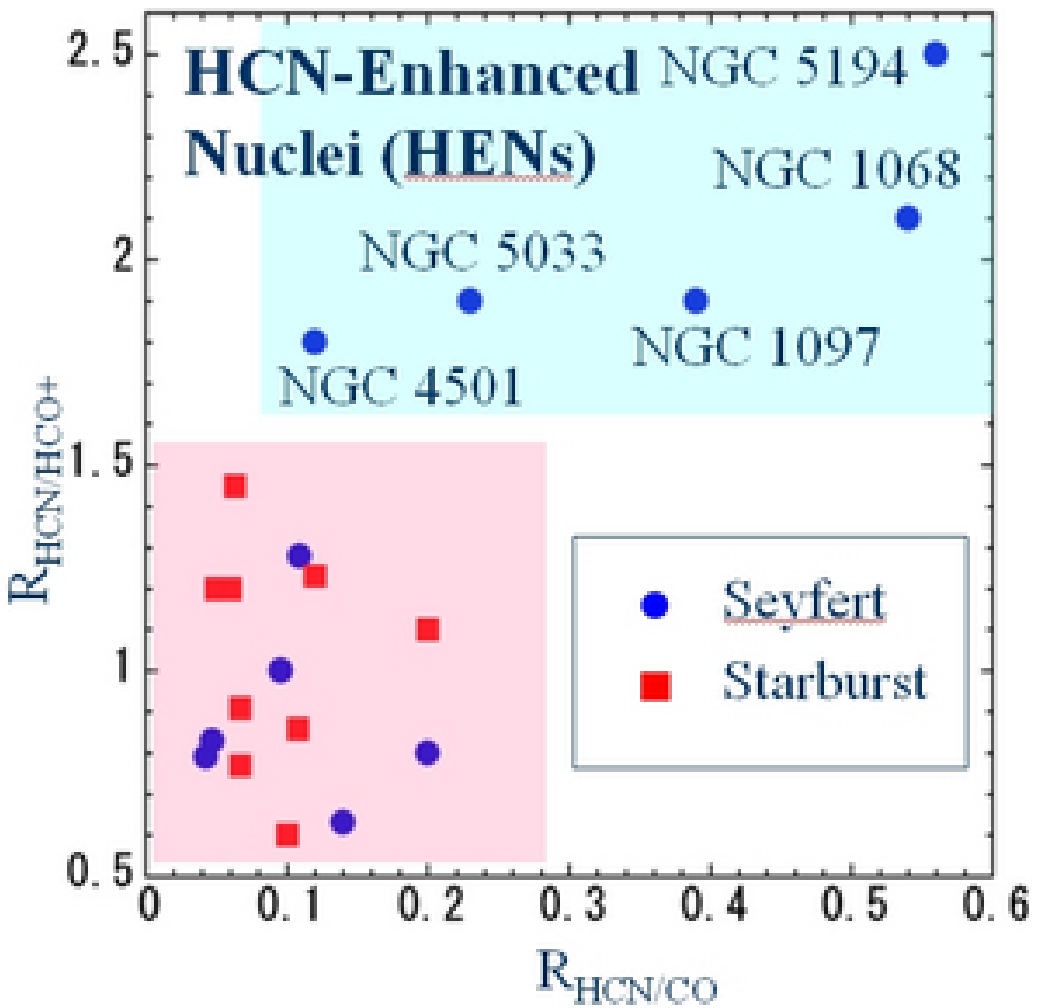} 
   \end{minipage}
   \begin{minipage}{.2\textwidth}
    \hfill
   \end{minipage}
   \begin{minipage}{.4\textwidth}   
{\bf Fig.1.}
The distributions of the observed $R_{\rm HCN/CO}$ and $R_{\rm HCN/HCO+}$ ratios 
at the center of a few 100 pc regions
of nearby Seyfert and starburst galaxies. 
Some of the Seyfert galaxies exhibit enhanced HCN($J$=1--0)
emission with respect to HCO$^+$($J$=1--0) and/or CO($J$=1--0) intensities (HENs). 
HENs are never observed among starburst galaxies.
\end{minipage}
\end{figure}

\begin{figure}
\includegraphics[width=11.5cm]{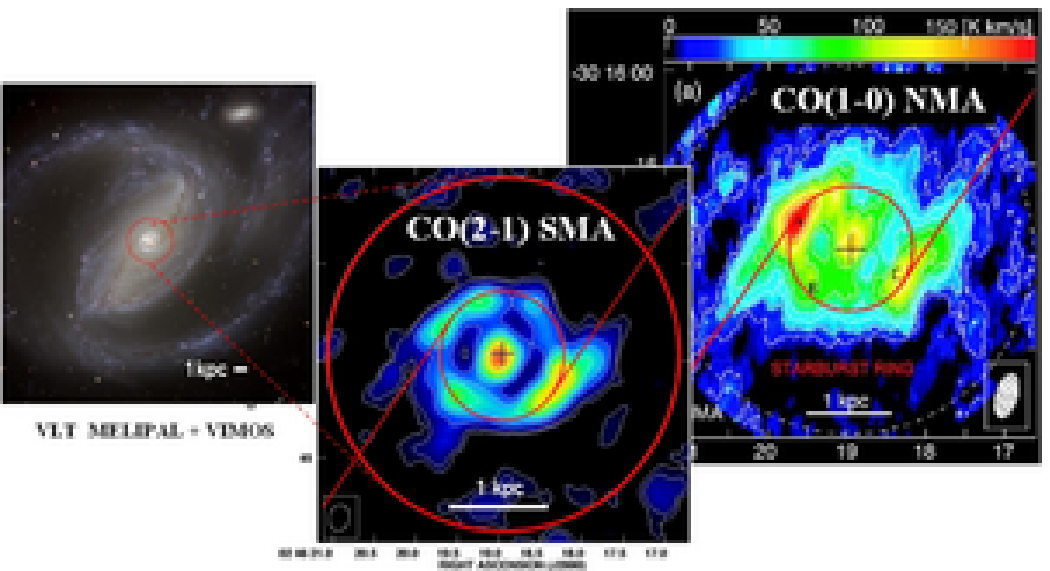} 
{\bf Fig.2.}
High-resolution integrated intensity images of CO($J$=2--1) 
and CO($J$=1--0) in the central kpc regions of NGC 1097
(Hsieh et al.\ 2007; Kohno et al.\ 2003). 
The CO($J$=1--0) emission is dominated by the circumnuclear starburst ring (indicated
by a red circle), whereas a strong peak emerges at the center in CO($J$=2--1).
\end{figure}

\subsection{CN($N$=1--0) in local starburst and Seyfert galaxies} 

We have also started a pilot survey of the CN($N$=1--0) emission toward nearby starburst and Seyfert
galaxies using NMA. The CN emission is another strong and important tracer of dense molecular gas
after HCN and HCO$^+$ (\cite{Aalto2002}; \cite{Riechers2007}), 
and CN/HCN integrated intensity ratios ($R_{\rm CN/HCN}$) are
expected to be enhanced significantly in the PDRs (\cite{Boger2005}) 
and XDRs (\cite{Lepp1996}; \cite{Meijerink2007}). One of our goals is to obtain spatially
resolved CN maps and to compare them with our existing HCN maps; this will allow us
to understand the relative importance of PDR and XDR
on the CN abundance.
The instantaneous bandwidth of the NMA spectrocorrelator 
(1024 MHz; \cite{Okumura2000}) 
enables us to observe 
$J$=$\frac{3}{2}-\frac{1}{2}$ and $J$=$\frac{1}{2}-\frac{1}{2}$ lines (consisting of
hyperfine structure lines) simultaneously.
Their ratios (hereafter $R_{(\frac{3}{2}-\frac{1}{2})/(\frac{1}{2}-\frac{1}{2})}$) 
can be used to estimate their opacities.

Our preliminary results are presented in Fig.~3.
All observed starburst galaxies 
(including NGC 3079, which hosts an AGN accompanied with a nuclear starburst) 
show an $R_{(\frac{3}{2}-\frac{1}{2})/(\frac{1}{2}-\frac{1}{2})}$ of $\sim$3 
(i.e., optically thin), implying that CN($N$=1--0) emission can be a useful measure
of mass or abundance of dense ISM.
On the other hand, 
$R_{(\frac{3}{2}-\frac{1}{2})/(\frac{1}{2}-\frac{1}{2})}$ was found to be close to unity 
(1.4 $\pm$ 0.15 within the central $r<3''$) at the center of NGC 1068,
suggesting that these lines are almost optically thick. 
This could be due to a significant increase in CN abundance
at the center of NGC 1068, a prototypical giant XDR (\cite{Usero2004}).

We then computed the $R_{\rm CN/HCN}$ values for two positions of NGC 1068, 
i.e., the center and the circumnuclear starburst region.
There was no significant difference in the $R_{\rm CN/HCN}$ values between
the nucleus and starburst region ($R_{\rm CN/HCN}$ was around 1.5 for both positions).
This could indicate that {\it both} CN and HCN abundances are significantly elevated 
in the XDR of NGC 1068.
Again, these results appear somewhat inconsistent with theoretical predictions 
(e.g., \cite{Meijerink2007}), 
and we do require further studies both observationally and theoretically.

\begin{figure}
\includegraphics[width=11.5cm]{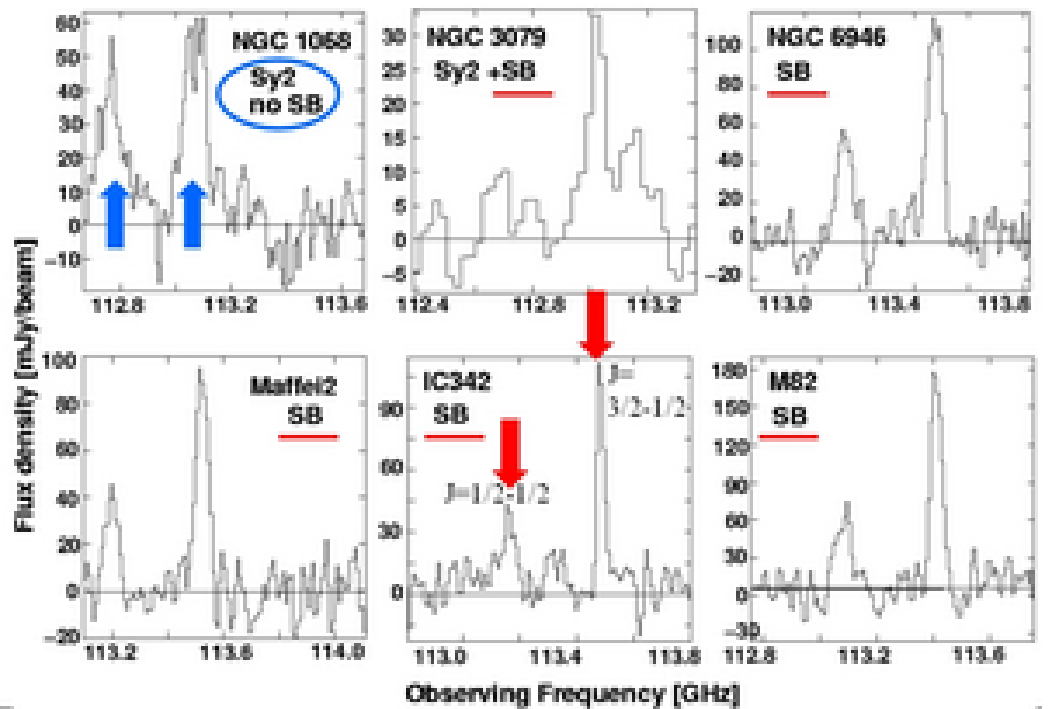} 
{\bf Fig.3.}
A part of CN($N$=1--0) spectra in the central regions of local starburst and
Seyfert galaxies taken with the NMA in the D-configuration
(the observing beam sizes are typically $\sim 6''$).
\end{figure}

\section{Tracing star-forming dense ISM in the disk regions of galaxies}

We have conducted CO($J$=3--2) imaging observations of nearby star-forming galaxies
(ADIoS project; \cite{Kohno2008b}; \cite{Muraoka2007}; \cite{Tosaki2007}) 
in order to depict star-forming dense molecular material
in the disk regions of galaxies, where high-density tracers such as HCN and HCO$^+$ molecules
are very weak. Even with the single pixel receiver of ASTE 345 GHz band, 
the excellent atmospheric conditions of Atacama, good performance of the antenna and receiver,
and efficient OTF mapping mode (\cite{Sawada2008}) enable us to rapidly produce 
wide area CO($J$=3--2) maps. Based on these data, a tight and linear connection 
between star formation rates (SFRs) and 
the amount of dense molecular gas traced by the CO($J$=3--2) emission is revealed
(Fig.~4).

\begin{figure}
\includegraphics[width=11.5cm]{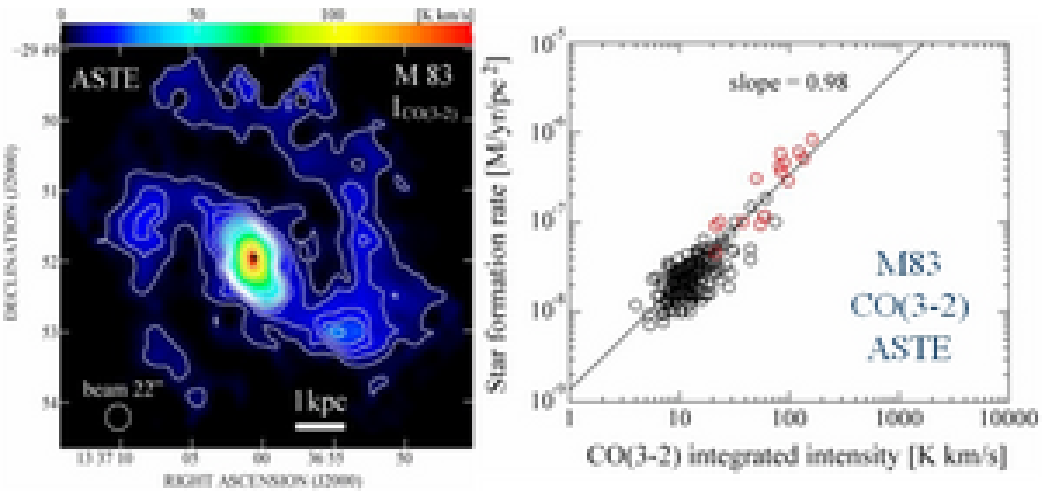} 
{\bf Fig.4.}
(Left) The CO($J$=3--2) integrated intensity map of the central $5' \times 5'$ region 
of the barred spiral galaxy M 83 (Muraoka \etal\ in prep).
(Right) The correlation diagram between the observed CO(3-2) integrated intensities
and surface densities of SFRs 
derived from extinction-corrected H$\alpha$ luminosities 
based on MIPS 24 $\mu$m data (\cite{Calzetti2007}). 
See also Komugi \etal\ (2007).
\end{figure}

\section{Tracing hidden violent star formation in the early Universe}

To uncover the hidden dusty massive star formation in the early Universe,
we have started a deep $\lambda$ 1.1 mm continuum imaging survey 
toward various blank fields and biased regions 
(such as proto-cluster forming regions around high-z radio galaxies, and so on) 
using the 144-pixel bolometer camera AzTEC (\cite{Wilson2008}) mounted on ASTE.
A typical map size is $12' \times 12'$ or wider, and a typical 1 $\sigma$ noise level
is $\sim$1 mJy or lower, allowing us to investigate
large scale distributions/structures of SMGs (i.e., $\sim$ a few 10 Mpc scales) 
with a sufficient sensitivity to detect hyper luminous IR galaxies 
(i.e., $L_{\rm IR} > 10^{13} L_\odot$) in the early Universe.

One of the deep images obtained is presented in Fig.~5. Our preliminary combined analysis
with AKARI 90 $\mu$m data suggests most of these sources are lying high-$z$ ($z>1$). 

In the Ly$\alpha$ selected proto-cluster region SSA 22, 
about a 390 arcmin$^2$ region was deeply observed. We found a very poor spatial correlation
between the detected 1.1 mm sources and the Ly$\alpha$ sources (LAEs and LABs), 
which was consistent with a previous finding in the TN J1338-1942 field (\cite{DeBreuck2004}).
In spite of such a poor overlap between SMGs and LAEs,
we found that the bright 1.1 mm sources are concentrated
around the density peak of LAEs, i.e., a possible excess of number density 
of SMGs around the LAEs' density peak can be observed in our data.
Further analysis, including the multiwavelengths
follow-up observations to obtain constraints on the source redshifts, is in progress.

\begin{figure}
   \begin{minipage}{.4\textwidth}
    \hfill 
\includegraphics[width=7cm]{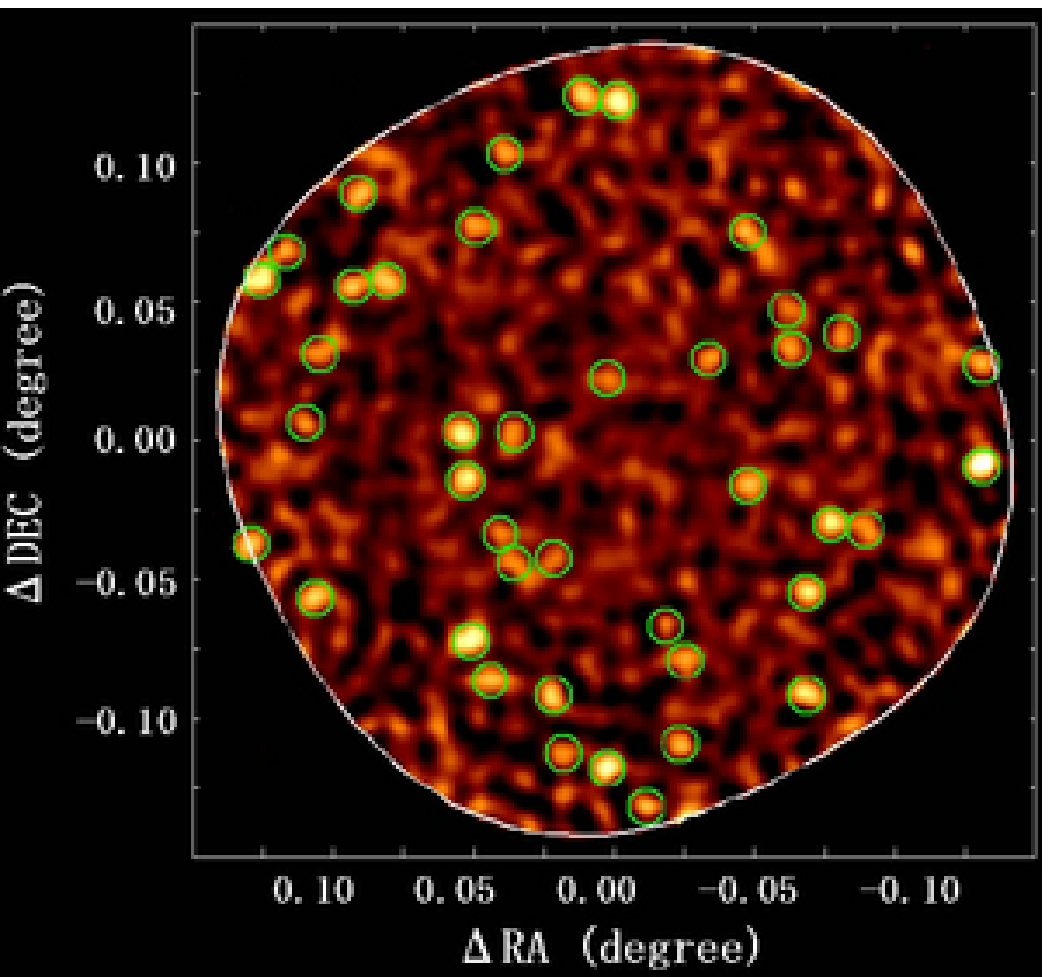} 
   \end{minipage}
   \begin{minipage}{.2\textwidth}
    \hfill
   \end{minipage}
   \begin{minipage}{.4\textwidth}   
{\bf Fig.5.}
An example of $\lambda$ 1.1 mm images taken with the AzTEC on ASTE
(Hatsukade et al., in prep.). 
A $\sim$200 arcmin$^2$ region was observed with a sensitivity
of $\sim$0.5 mJy toward a part of AKARI Deep Field South (ADF-S), 
where an extensive infrared deep survey
has been conducted using the AKARI satellite (\cite{Matsuhara2006}).
More than 20 high significance (S/N $>$ 4) sources have been detected.  
The beam size of AzTEC/ASTE is $\sim 28''$.
\end{minipage}
\end{figure}


\end{document}